\documentclass{elsart}
\usepackage{amsfonts}
\usepackage{amsmath}
\usepackage{amssymb}
\usepackage{epsfig}
\newcommand{\rme}{\mathrm{e}}
\newcommand{\cA}{\mathcal A}
\newcommand{\cS}{\mathcal S}
\newcommand{\ie}{{i.e.\ }} 
\begin{document}
\begin{frontmatter}
\title{Controlling chaos in area-preserving maps}

\author[cpt]{C. Chandre},
\author[cpt]{M. Vittot},
\author[piim]{Y. Elskens},
\author[cpt]{G. Ciraolo}, 
and \author[inaf]{M. Pettini}
\address[cpt]{Centre de
Physique Th\'eorique - CNRS, Luminy - Case 907, F-13288 Marseille
cedex 09, France}
\address[piim]{UMR 6633 CNRS -- Universit\'e de Provence,
Centre Scientifique de St J\'er\^ome, Case 321, F-13397 Marseille cedex 20, France}
\address[inaf]{Istituto Nazionale di Astrofisica, Largo Enrico Fermi 5, I-50125 Firenze, Italy,
I.N.F.M. UdR Firenze and I.N.F.N. Sezione di Firenze}

\begin{abstract}
We describe a method of control of chaos that occurs in area-preserving maps. This method is based on small modifications of the original map by addition of a small control term. We apply this control technique to the standard map and to the tokamap.
\end{abstract}
\end{frontmatter}

%=========================================================== 
\section{Introduction}
%=========================================================== 

Chaos that arises naturally in Hamiltonian systems with mixed phase space is a source of unstable behaviors when resonances overlap. Achieving the control of these systems by reducing or even suppressing chaos is a long standing  and crucial problem in many branches of physics. Among the various methods which have been designed to control Hamiltonian chaos, two different frameworks must be distinguished~: Hamiltonian flows (continuous time)~\cite{flows,flows2}, and symplectic maps (discrete time)~\cite{upo1,upo2,upo2b,upo3,upo4,emb2,modPulse1,sym1,modPulse2,mezi}.

A method for controlling Hamiltonian flows has been proposed in Refs.~\cite{guido1,guido2,michel,guido4}. Its aim was to find a small control term $f$ for the perturbed Hamiltonian $H=H_0+\varepsilon V$ where $H_0$ is integrable, in order to have a more regular dynamics for the controlled Hamiltonian $H_0+\varepsilon V+f$. An explicit algorithm to construct a control term of order $\varepsilon^2$ making the controlled Hamiltonian integrable was devised. The main advantage of this approach to control is the robustness of the control term $f$, which means that one can tailor another control term $\tilde{f}$ which is close to $f$ such that $(i)$ the controlled Hamiltonian $H_0+\varepsilon V+\tilde{f}$ is still more regular than $H_0+\varepsilon V$ and $(ii)$ this control terms satisfies some specific requirements of the experiment like a partial knowledge of the potential $V$~\cite{guido2} or the localization of the control in a phase space region~\cite{guido4}. Let us stress that this method differs from other methods by the fact that the controlled dynamics is Hamiltonian~: This makes it relevant to the control of inherently Hamiltonian systems~\cite{meiss} such as beams of charged test particles in electrostatic waves~\cite{beam,beam2}, two-dimensional Euler flows~\cite{otti} or the geometry of magnetic field lines~\cite{toka1,toka2}. 

In this article, the problem we address is to set up a control theory in a similar way for symplectic maps.  
We consider two-dimensional symplectic maps $(A,\varphi)\mapsto (A',\varphi')=F(A,\varphi)$ on the cylinder ${\mathbb R}\times {\mathbb T}$ which are $\varepsilon$-close to integrability. Our aim is to find a small control term $f$ such that the controlled symplectic map $F+f$ is more regular (closer to integrability) than $F$. In what follows, we develop algorithms for finding control terms of order $\varepsilon^2$. 

We describe two approaches to the control of symplectic maps~: One approach (in section~\ref{sec2}) is designed for maps
which are generated by Lie transforms. We show that it is possible to obtain an explicit expression for a formal control term which makes the controlled map integrable. This approach extends in an obvious way to $2N$-dimensional symplectic maps.
However, since this approach is difficult to handle in practice, we develop a second one (in section~\ref{sec3}) for maps which are written in mixed coordinates (\ie half implicitly, half explicitly). This second method is used for practical purposes since numerous area-preserving maps are written in such form. 

In section~\ref{sec4} we apply our control method to two examples~: the standard map to benchmark our method, and the tokamap, which imposes an additional restriction on the type of control. We compute analytically the control term and show numerically the phase portrait of these maps and the associated controlled maps.
As the formal expansion may generate undesirable side-effects in phase space, we also discuss suitable modifications of the control terms. 

%=========================================================== 
\section{Maps expressed as Lie transforms}
%=========================================================== 
\label{sec2}

Let $\mathcal A$ be the vector space
of $C^{\infty}$ real functions defined on the
phase space ${\mathbb R}\times {\mathbb T}$. 
For $H\in \mathcal A$, let $\{H\}$ be the linear operator
acting on $\mathcal A$  such that
$$
\{H\}H^{\prime}=\{H,H^{\prime}\},
$$
for any $H^{\prime}\in \mathcal A$, where $\{\cdot~,\cdot\}$ is the Poisson bracket. 
Hence $\mathcal A$ is a Lie algebra. For two-dimensional maps written in action-angle coordinates, the linear operator $\{H\}$ writes
$$
\{H\}=\frac{\partial H}{\partial A}\frac{\partial ~ }{\partial \varphi}-\frac{\partial H}{\partial \varphi}\frac{\partial ~ }{\partial A}.
$$
We consider area-preserving maps generated by a Lie transform
\begin{equation}
\label{eqn:map}
\left(\begin{array}{c} A' \\ \varphi'\end{array}\right)={\mathrm e}^{\{H\}}{\mathrm e}^{\{V\}}
\left(\begin{array}{c} A \\ \varphi\end{array}\right),
\end{equation}
where $H$ and $V$ are two elements of the algebra $\mathcal A$ and the linear operator ${\mathrm e}^{\{H\}}$ is defined by the expansion
$$
{\mathrm e}^{\{H\}}=\sum_{n=0}^{\infty}\frac{\{H\}^n}{n!}.
$$
In what follows, we will use two warped additions, one for linear operators acting on ${\mathcal A}$ (elements of $L({\mathcal A})$) and another one for elements of $\mathcal A$~: First, we define the warped addition $\oplus$ between two elements $\alpha$ and $\beta $ of $L({\mathcal A})$ as
\begin{equation}
\label{eqn:defoplus}
{\mathrm e}^{\alpha \oplus \beta}={\mathrm e}^{\alpha}{\mathrm e}^{\beta}.
\end{equation}
An explicit series for $\alpha \oplus \beta$ is given by the Baker-Campbell-Hausdorff formula~\cite{BCH}. Specializing to $\alpha=\{H\}$ and $\beta=\{V\}$ for $H,V\in{\mathcal A}$, one can see that $\{H\}\oplus \{V\}$ is also a Lie operator~: For instance, the first terms of $\{H\}\oplus \{V\}$ are
\begin{eqnarray*}
\{H\}\oplus \{V\}&=&\{H\}+ \{V\}+\frac{1}{2} \left( \{H\} \{V\}-\{V\}\{H\}\right)+\cdots, \\
                 &=&\{H + V+\frac{1}{2} \{H\} V+\cdots \}, 
\end{eqnarray*}
where the second equation follows from the Jacobi identity.
We associate with $\{H\}\oplus \{V\}$ an element of $\mathcal A$ which we denote $H\boxplus V$ if it exists (it does, e.g., when $H$ and $V$ are analytic in $A$ and $\varphi$, and $V$ small enough)~:
\begin{equation}
\label{eqn:defboxplus}
\{H\} \oplus \{V\}=\{ H\boxplus V\},
\end{equation}
or equivalently 
$$
{\mathrm e}^{\{H \boxplus V\}}={\mathrm e}^{\{H\}}{\mathrm e}^{\{V\}},
$$
where the first terms of the expansion of $H\boxplus V$ are
$$
H\boxplus V=H + V+\frac{1}{2} \{H\} V+\cdots.
$$
Both warped additions $\oplus$ and $\boxplus$ are associative but not commutative. Also, we have the following two properties~: $W=H\boxplus V$ implies $H=W\boxplus (-V)$, and $(-H)\boxplus(-V)=-(V\boxplus H)$.
The map given by Eq.~(\ref{eqn:map}) will be called the map generated by $H \boxplus V$.

For instance, if $H$ depends only on the action $A$, and $V$ only on the angle $\varphi$, the map obtained from Eq.~(\ref{eqn:map}) is the composition of the following two maps
$$
\left(\begin{array}{c} A' \\ \varphi'\end{array}\right)={\mathrm e}^{-\partial_\varphi V(\varphi)\partial_A}
\left(\begin{array}{c} A \\ \varphi\end{array}\right)=\left(\begin{array}{c} A-\partial_\varphi V (\varphi) \\ \varphi\end{array}\right),
$$
and
$$
\left(\begin{array}{c} A'' \\ \varphi''\end{array}\right)={\mathrm e}^{\omega(A)\partial_\varphi}
\left(\begin{array}{c} A' \\ \varphi'\end{array}\right)=\left(\begin{array}{c} A' \\ \varphi'+\omega(A')\end{array}\right),
$$
where $\omega(A)=\partial_A H$ and $\partial_A$ and $\partial_\varphi$ denote the partial derivatives with respect to $A$ and $\varphi$. Thus
\begin{eqnarray*}
	&& A''=A-\partial_\varphi V(\varphi),\\
	&& \varphi''=\varphi+\omega(A-\partial_\varphi V(\varphi)).
\end{eqnarray*}
When $V$ depends also on the action $A$, the map is much more difficult to write explicitly.

From the operator $\{H\}$, we define the linear operator
$$
{\mathcal H}=1-{\mathrm e}^{-\{H\}}.
$$
This operator $\mathcal H$ is not invertible~: for
instance $ { \mathcal H} H = 0 $. 
Hence we consider a pseudo-inverse of $\mathcal H$ (if it exists), \ie
a linear operator $\Gamma$ on $\mathcal A$ such that
\begin{equation}
\label{eqn:gamma}
{\mathcal H}^2\Gamma ={\mathcal H}.
\end{equation}
From this operator, we define the operators $\mathcal R$ and $\mathcal N$ as
\begin{eqnarray*}
	&& {\mathcal N}={\mathcal H}\Gamma,\\
	&& {\mathcal R}=1-{\mathcal N}.
\end{eqnarray*}
Note that ${\mathcal H}{\mathcal R}=0$. If $\Gamma$ commutes with ${\mathcal H}$, the operators $\mathcal R$ and $\mathcal N$ are projectors. 

From now on, we assume that $H$ depends only on the action $A$. Then the pseudo-inverse $\Gamma$ exists formally. The operators $\{H\}$, ${\mathcal H}$, $\Gamma$, ${\mathcal R}$ and ${\mathcal N}$ act on a function $U(A,\varphi)=\sum_{k\in{\mathbb Z}} U_k(A){\mathrm e}^{ik\varphi}$ as 
\begin{eqnarray*}
	&& \{H\} U=\sum_k ik\omega(A) U_k(A) {\mathrm e}^{ik\varphi},\\
	&& {\mathcal H} U=\sum_{k}(1-{\mathrm e}^{-i\omega(A)k}) U_k(A){\mathrm e}^{ik\varphi},\\
	&&\Gamma U=\sum_{k \mbox{ s.t. } \omega(A)k\notin 2\pi{\mathbb Z}} \frac{U_k(A)}{1-{\mathrm e}^{-i\omega(A)k}}{\mathrm e}^{ik\varphi},\\
	&& {\mathcal R}U=\sum_{k \mbox{ s.t. } \omega(A)k\in 2\pi{\mathbb Z}} U_k(A) {\mathrm e}^{ik\varphi},\\
	&& {\mathcal N}U=\sum_{k \mbox{ s.t. } \omega(A)k\notin 2\pi{\mathbb Z}} U_k(A) {\mathrm e}^{ik\varphi}.
\end{eqnarray*}

\begin{prop} There exists a control term $f$ of order $V^2$ such that the map generated by $H\boxplus V\boxplus f$ is canonically conjugate to the map generated by $H\boxplus{\mathcal R}V$. The control term is given by the expression \begin{equation} \label{eqn:ctH} f=(-V)\boxplus ({\mathcal N}V-\Gamma V)\boxplus {\mathcal R}V\boxplus \Gamma V \end{equation} and the conjugation is generated by $\Gamma V$. \end{prop}

In other words, the controlled map is
  \begin{equation}
\label{eqn:Cmap}
\left(\begin{array}{c} A' \\ \varphi'\end{array}\right)={\mathrm 
e}^{\{H\}}{\mathrm e}^{\{V\}}{\mathrm e}^{\{f\}} \left(\begin{array}{c} A \\ \varphi\end{array}\right), \end{equation} where ${\mathrm e}^{\{f\} }={\mathrm e}^{-\{V\}}{\mathrm e}^{\{{\mathcal N} 
V-\Gamma V\}}{\mathrm e}^{\{{\mathcal R}V\}} {\mathrm e}^{\{\Gamma V\} }$.

{\em Proof:} The conjugation equation $$ {\mathrm e}^{\{\Gamma V\}} \left(H\boxplus V\boxplus f\right)=H\boxplus {\mathcal R}V, $$ has the solution \begin{equation} \label{eqn:eqct}
   f = (-V) \boxplus (-H) \boxplus
       {\mathrm e}^{-\{\Gamma V\}}(H\boxplus{\mathcal R}V). \end{equation} In Appendix~\ref{app:opl} we prove the identity \begin{equation} \label{eqn:lem} {\mathrm e}^{\{ U\}} W=U\boxplus W\boxplus(-U), \end{equation} for functions $U,W \in {\mathcal A}$. Applying Eq.~(\ref{eqn:lem}) twice in Eq.~(\ref{eqn:eqct}) yields \begin{eqnarray*}
   f &=& (-V)\boxplus (-H)\boxplus (-\Gamma V)\boxplus H\boxplus {\mathcal 
R}V\boxplus \Gamma V
   \cr
   &=& (-V)\boxplus( -{\mathrm e}^{-\{H\}} \Gamma V)\boxplus {\mathcal 
R}V\boxplus \Gamma V ,
\end{eqnarray*}
which reduces to Eq.~(\ref{eqn:ctH}) as ${\mathrm e}^{-\{H\}} 
\Gamma=\Gamma-{\mathcal N}$ by the definition of ${\mathcal N}$. \hfill 
$\Box$ \\

Note that the map generated by $H\boxplus {\mathcal R}V$ is not integrable in general (since ${\mathcal R}V$ does depend on the angle $\varphi$). However, we point out two specific cases where it is integrable: The first one is when the frequency $\omega/(2\pi)$ is irrational and the second one is in the resonant case when $V$ does not contain any mode $k\not= 0$ such that $k\omega\in {\mathbb Z}$.
In these two cases, ${\mathcal R} V$ depends only on the action $A$. Therefore, the controlled map generated by $H\boxplus V\boxplus f$ is integrable since it is canonically conjugate to $H\boxplus {\mathcal R}V=H+ {\mathcal R}V $.

The Baker-Campbell-Hausdorff formula yields the expansion of the control term $f$~: Its term linear in $V$ which is $-V+({\mathcal N}V-\Gamma V)+{\mathcal R}V+\Gamma V$, vanishes, so that its dominant term is of order $V^2$~:
\begin{equation}
\label{eqn:expf}
	f=-\{\Gamma V\}\frac{{\mathcal R}+1}{2} V+\frac{1}{2}\{V\} {\mathcal R}V +O(V^3).
\end{equation}
Note that in the continuous case, the dominant term of the control term was~\cite{guido1,michel}
$$
f_2=-\{\Gamma V\}\frac{{\mathcal R}+1}{2} V.
$$ 
By introducing the time step $\tau$ between two iterations of the map, we define the operator 
$$
{\mathcal H}_\tau=\frac{1-{\mathrm e}^{-\tau \{H\}}}{\tau},
$$
and $\Gamma_\tau$ as the pseudo-inverse of ${\mathcal H}_\tau$ by Eq.~(\ref{eqn:gamma}). The projectors ${\mathcal R}_\tau$ and ${\mathcal N}_\tau$ are defined accordingly.
The control term is given by
$$
f_\tau=\tau^{-1}\left[ (-\tau V)\boxplus (\tau {\mathcal N}_\tau V- \Gamma_\tau V)\boxplus \tau {\mathcal R}_\tau V\boxplus \Gamma_\tau V \right].
$$ 
The dominant terms of $f_\tau$ are given by 
$$
f_\tau =-\{\Gamma_\tau V\}\frac{{\mathcal R}_\tau+1}{2} V+\frac{\tau}{2}\{V\} {\mathcal R}_\tau V +O(V^3).
$$
Therefore when $\tau$ tends to zero, the dominant term of the control term for symplectic maps tends to the one of the time-continuous Hamiltonian systems. 

{\em Remark~:} If we choose the controlled map to be generated by $H\boxplus \tilde{f} \boxplus V$, then the control term $\tilde{f}$ is defined from the control term $f$ as $\tilde{f}={\mathrm e}^{\{V\}}f$. The dominant term of $\tilde{f}$ coincides with the one of $f$ (higher order terms do not coincide in general). If the controlled map is chosen to be generated by $\hat{f}\boxplus H\boxplus V$, the control term $\hat{f}$ is $\hat{f}={\mathrm e}^{\{H\}} \tilde{f}$.

%=========================================================== 
\section{Maps in mixed coordinates}
%=========================================================== 
\label{sec3}
 
We consider area-preserving maps obtained from a generating function of the form
$$
S(A',\varphi)=A'\varphi+H(A')+\varepsilon V(A',\varphi).
$$
In this section, we kept $\varepsilon$ for bookkeeping purposes. 
The map reads
\begin{eqnarray}
	&& A=A'+\varepsilon \partial_\varphi V(A',\varphi),\label{eqn:11a}\\
	&& \varphi'=\varphi+\omega(A')+\varepsilon \partial_A V(A',\varphi),\label{eqn:11b}
\end{eqnarray}
where $\omega(A)=\partial_A H$. Here $\partial_A V(A',\varphi)$ denotes the partial derivative of $V$ with respect to the action (first variable) and  $\partial_\varphi V(A',\varphi)$ denotes the partial derivative of $V$ with respect to the angle (second variable).

Note that for $\varepsilon=0$, the map (\ref{eqn:11a})-(\ref{eqn:11b}) is integrable and identical to the map generated by the Lie transform $\rme^{\{H\}}$. Thus we define the linear operators $\mathcal H$, $\Gamma$, $\mathcal R$ and $\mathcal N$ as above.

\begin{figure}
\epsfig{file=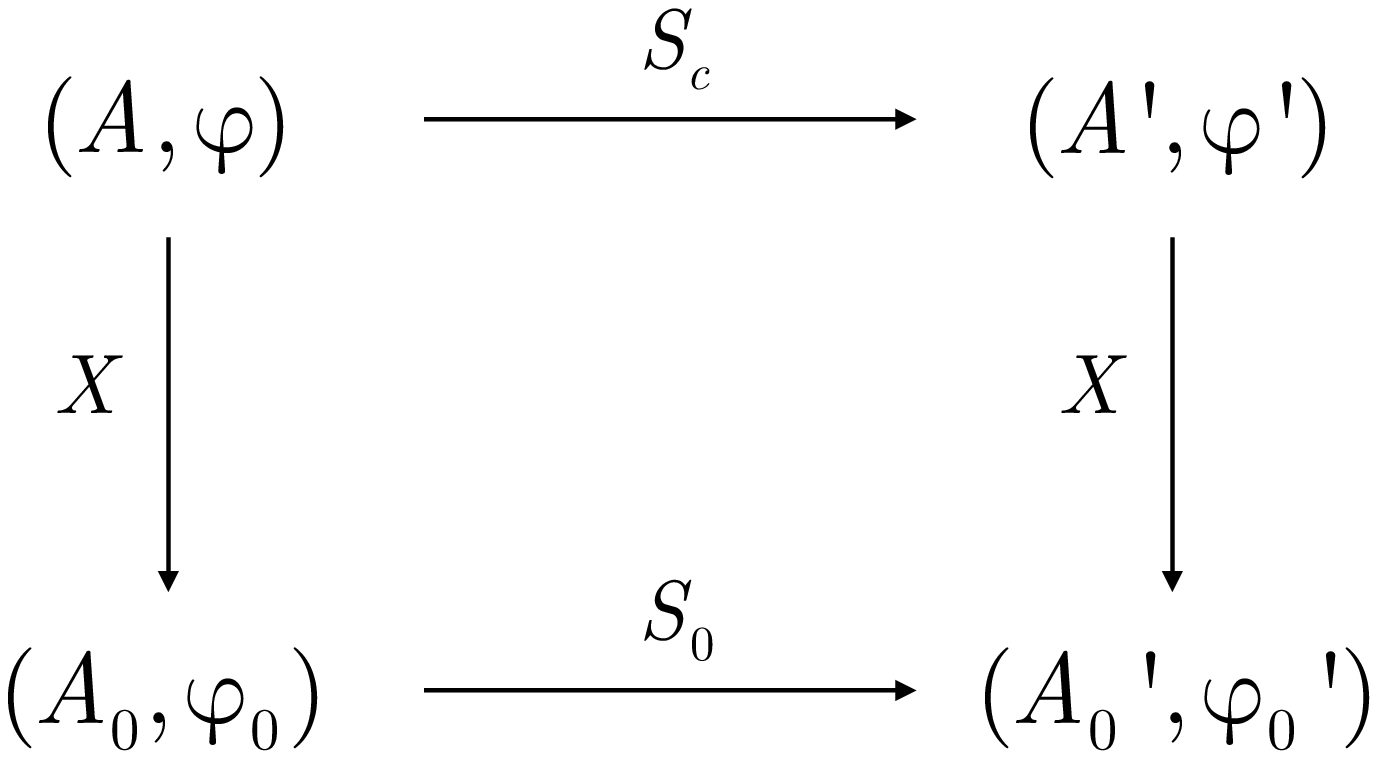,height=5cm,width=9.5cm}
\caption{Diagram of the generating functions for the canonical changes of variables.}
\label{fig:map}
\end{figure}

Our aim is to modify the generating function with a control term of order $\varepsilon^2$ such that the controlled map is closer to integrability than the original map. We consider the controlled generating function
$$
S_{\mathrm{c}}(A',\varphi)=A'\varphi+H(A')+\varepsilon V(A',\varphi)+\varepsilon^2 f(A',\varphi).
$$
To construct the control term $f$, we perform a near-identity canonical transformation with generating function $X(A_0,\varphi)=A_0\varphi+\varepsilon \chi (A_0,\varphi)$ which maps $(A,\varphi)$ into $(A_0,\varphi_0)$ and $(A',\varphi')$ into $(A_0',\varphi_0')$ using the same function $\chi$. The scheme for the change of coordinates of the original map is depicted on Fig.~\ref{fig:map}.
The canonical change of coordinates maps the system obtained by the generating function $S_{\mathrm{c}}$ into a system obtained by the generating function $S_0$~:
\begin{eqnarray}
\label{eqn:s0}
    S_0(A'_0,\varphi_0)
      &=& A'_0\varphi_0+H(A'_0)+\varepsilon z(A'_0,\varphi_0)\nonumber \\
    && + \varepsilon^2\Bigl[- \partial_\varphi z (A'_0,\varphi_0) \partial_A\chi(A'_0,\varphi_0)
             + \partial_\varphi\chi(A'_0,\varphi_0+\omega(A'_0))\partial_A V(A'_0,\varphi_0)\nonumber\\
    && +\frac{1}{2}\omega'(A'_0)\bigl(\partial_\varphi\chi(A'_0,\varphi_0+\omega(A'_0))\bigr)^2+f(A'_0,\varphi_0) \Bigr]
         + O(\varepsilon^3),
\end{eqnarray}
where 
\begin{equation}
\label{eqn:defz}
z(A'_0,\varphi_0)=V(A'_0,\varphi_0)+\chi(A'_0,\varphi_0+\omega(A'_0))-\chi(A'_0,\varphi_0).
\end{equation}
We refer to Appendix~\ref{app:1} for the detailed computations.

Now we choose the generating function $\chi$ such that the order $\varepsilon$ in Eq.~(\ref{eqn:s0}) vanishes. To determine $\chi$, we expand $V$ as $V={\mathcal R}V+{\mathcal N}V$. We require that~:
$$
\chi(A,\varphi)-\chi(A,\varphi+\omega(A))={\mathcal N}V(A,\varphi).
$$
By expanding $V$ in Fourier series, \ie $V(A,\varphi)=\sum_{k\in {\mathbb Z}} V_k(A){\mathrm e}^{ik\varphi}$, this reads
\begin{equation}
\label{eqn:chi}
\chi(A,\varphi)=\sum_{k \mbox{ s.t. } \omega(A)k\notin 2\pi{\mathbb Z}} \frac{V_k(A)}{1-{\mathrm e}^{i\omega(A)k}} {\mathrm e}^{ik\varphi},
\end{equation}
which means that $\chi(A,\varphi+\omega(A))=-\Gamma V(A,\varphi)$. As $\chi(A,\varphi+\omega(A))={\mathrm e}^{\{H\}} \chi(A,\varphi)$, we find 
\begin{equation}
\label{eqn:chimix}
\chi={\mathcal N}V-\Gamma V.
\end{equation}
The control term is chosen such that the order $\varepsilon^2$ of $S_0$ given by Eq.~(\ref{eqn:s0}) vanishes. Its expression 
\begin{equation}
\label{eqn:hmix}
f(A,\varphi)=\left(\partial_A V\right) \partial_\varphi \Gamma V-\frac{\partial_A \omega}{2}\left(\partial_\varphi \Gamma V \right)^2 +\left(\partial_\varphi {\mathcal R}V\right) \partial_A ({\mathcal N}V- \Gamma V),
\end{equation}
represents the dominant term (of order $\varepsilon^2$) of a control term that would cancel all the terms of higher orders in $S_0$. Using this control term, the controlled map obtained from the generating function $S_{\mathrm{c}}$ is conjugate to the map generated by
$$
S_0(A',\varphi)=A'\varphi+H(A')+\varepsilon {\mathcal R}V+O(\varepsilon^3).
$$ 
As in the preceding section, in the two cases where $\omega$ is irrational or $V$ does not contain resonant modes, the map generated by $S_0$ is $\varepsilon^3$-close to integrability.

%=========================================================== 
\section{Numerical examples}
%=========================================================== 
\label{sec4}
%=========================================================== 
\subsection{Application to the standard map}
\begin{figure}
\epsfig{file=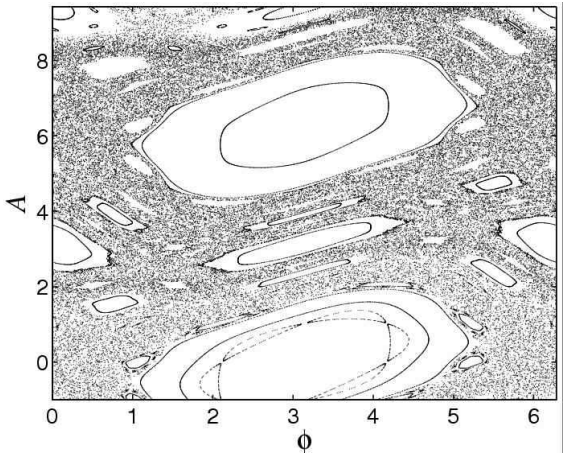,width=7.5cm,height=6.0cm}
\caption{Phase portrait of the standard map $\cS$ for $\varepsilon=1.2$.}
\label{fig:sm1}
\end{figure}

%=========================================================== 
The standard map $\cS$ is 
\begin{eqnarray*}
	&& A'=A+\varepsilon \sin \varphi,\\
	&& \varphi'=\varphi+A' \mbox{ mod } 2\pi. 
\end{eqnarray*}
A phase portrait of this map for $\varepsilon=1.2$ is given in Fig.~\ref{fig:sm1}. There are no Kolmogorov-Arnold-Moser (KAM) tori (acting as barriers in phase space) at this value of $\varepsilon$ (and higher). The critical value of the parameter $\varepsilon$ for which all KAM tori are broken is $\varepsilon_{\mathrm{std}}\approx 0.9716$.

\begin{figure}
\epsfig{file=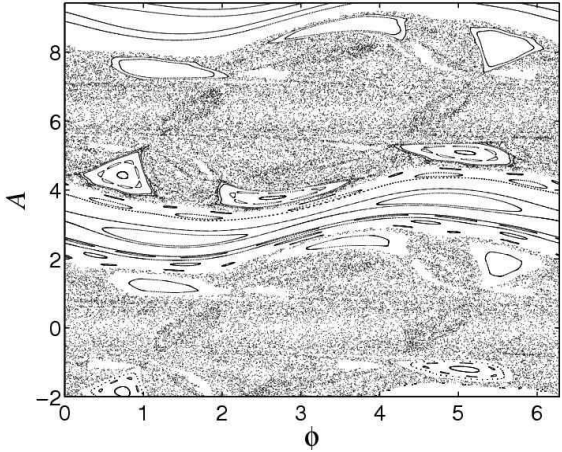,width=7.5cm,height=6.0cm}
\caption{Phase portrait of the controlled standard map $\cS_{\mathrm{mix},2}$ with the control term~(\ref{eqn:smctt}) for $\varepsilon=1.2$.}
\label{fig:sm3}
\end{figure}

The standard map is obtained from the generating function in mixed coordinates
$$
S(A',\varphi)=A'\varphi+\frac{A^{\prime 2}}{2}+\varepsilon \cos\varphi,
$$ 
\ie $V(A,\varphi)=\varepsilon\cos\varphi$ and $H(A)=A^2/2$. The generating function $\chi$ given by Eq.~(\ref{eqn:chi}) is thus
$$
\chi(A,\varphi)=-\varepsilon \frac{\sin(\varphi-A/2)}{2\sin(A/2)}.
$$
The dominant control term given by Eq.~(\ref{eqn:hmix}) is 
\begin{equation}
\label{eqn:smctt}
f_{\mathrm{mix},2}(A,\varphi)=-\varepsilon^2 \frac{\cos^2(\varphi+A/2)}{8\sin^2(A/2)},
\end{equation}
and the resulting map $\cS_{\mathrm{mix},2}$ generates the phase portrait displayed on Fig.~\ref{fig:sm3}. 

Note that besides the set of invariant tori which have been created in between the two primary resonances, the modification of phase space is significant. This comes from the fact that the control term $f_{\mathrm{mix},2}$ does not induce a small modification of the standard map~: actually, it is unbounded when $A$ approaches a primary resonance, \ie when $A\in 2\pi {\mathbb Z}$. 

To obtain a smaller control term, we will localize it near a region where $A$ is close to $\pi$ modulo $2\pi$. This corresponds to the centered resonance approximation~\cite{cra1,cra2,beam2}. If one keeps only the leading term, the control term becomes independent of $A$ (which may also be easier to implement in practice)~:
\begin{equation}
\label{eqn:ctsma}
f_{\mathrm{a}}(\varphi)=-\frac{\varepsilon^2}{8}\sin^2\varphi.
\end{equation}
The controlled standard map $\cS_{\mathrm{c,a}}$ generated by $f_{\mathrm{a}}$ is
\begin{eqnarray*}
	&& A'=A+\varepsilon \sin \varphi +\frac{\varepsilon^2}{8}\sin 2\varphi,\\
	&& \varphi'=\varphi+A' \mbox{ mod } 2\pi. 
\end{eqnarray*}

\begin{figure}
\epsfig{file=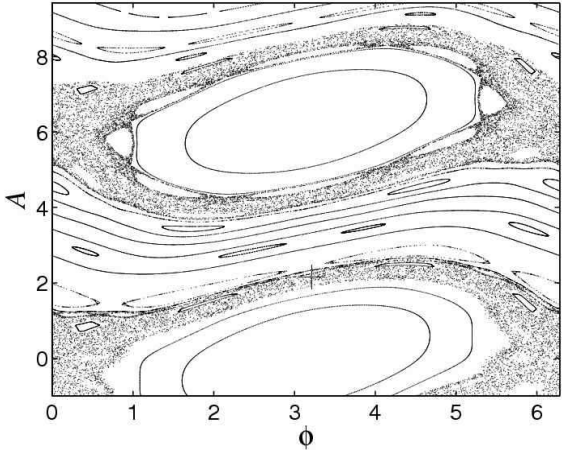,width=7.5cm,height=6.0cm}
\caption{Phase portrait of the controlled standard map ${\mathcal S}_{\mathrm c,a}$ with the approximate control term~(\ref{eqn:ctsma}) for $\varepsilon=1.2$.}
\label{fig:sm2}
\end{figure}

The phase portrait of $\cS_{\mathrm{c,a}}$ is displayed on Fig.~\ref{fig:sm2}. It clearly shows that adding the control term $f_{\mathrm{a}}$ created a lot of invariant tori in the region near $A=\pi$ whereas the structures near the primary islands are essentially preserved. 

If we consider higher order terms in the expansion near $A=\pi \mbox{ mod } 2\pi$, the control term becomes
\begin{equation}
\label{eqn:ctsmb}
f_{\mathrm{b}}(A,\varphi)=-\frac{\varepsilon^2}{8}\left( \sin^2\varphi-\frac{1}{2}\sin 2\varphi \sin A+\cos^2\varphi \cos^2(A/2)\right) +O(\cos^3(A/2)).
\end{equation}
If we neglect the cubic order, the controlled map $\cS_{\mathrm{c,b}}$ becomes
\begin{eqnarray*}
	&& A=A'-\varepsilon\sin\varphi -\frac{\varepsilon^2}{8}(\sin 2\varphi \sin^2(A'/2)-\cos 2\varphi \sin A'),\\
	&& \varphi'=\varphi +A'+\frac{\varepsilon^2}{16}(\sin 2\varphi \cos A' +\cos^2\varphi \sin A').
\end{eqnarray*}
The phase portrait of $\cS_{\mathrm{c,b}}$ is very similar to the one of $\cS_{\mathrm{c,a}}$ shown on Fig.~\ref{fig:sm2}. In addition to the set of invariant tori created in the region near $A=\pi \mbox{ mod } 2\pi$ in a very similar way as for $\cS_{\mathrm{c,a}}$, it is worth noting that the invariant tori of $\cS_{\mathrm{c,b}}$ are more robust than the ones of $\cS_{\mathrm{c,a}}$ (\ie they survive to slightly higher values of $\varepsilon$). More precisely, $\cS_{\mathrm{c,a}}$ has invariant tori up to $\varepsilon_{\mathrm a} \approx 1.68$, whereas invariant tori of $\cS_{\mathrm{c,b}}$ survive up to $\varepsilon_{\mathrm b} \approx 1.81$. Note that the controlled map $\cS_{\mathrm{mix},2}$ has invariant tori up to $\varepsilon_{\mathrm{mix},2} \approx 1.66$ while the standard map has invariant tori up to $\varepsilon_{\mathrm{std}} \approx 0.97$. These values were obtained using Laskar's frequency map analysis~\cite{lask,lask2,lask3,lask4}.  

In order to compare the control term obtained with mixed coordinates and the one using the Lie transform, note that the standard map is obtained by a Lie transform using $H(A)=A^2/2$ and $V(\varphi)=\varepsilon\cos\varphi$. For $A\notin 2\pi {\mathbb Z}$, we notice that ${\mathcal R}V=0$.  The action of the operator $\Gamma $ on $V$ becomes
$$
\Gamma V=\varepsilon \frac{\sin(\varphi+A/2)}{2\sin(A/2)}.
$$
The dominant order of the control term given by Eq.~(\ref{eqn:expf}) becomes
$$
f_{\mathrm{Lie},2}=-\frac{\varepsilon^2}{8} \left(\frac{\sin\varphi}{\sin(A/2)}\right)^2.
$$

Since the controlled map is difficult to explicit analytically, we consider the simplified case where we consider only the region near $A=\pi$. The approximation of the control term becomes
$$
f_{\mathrm{Lie,a}}=-\frac{\varepsilon^2}{8}\sin^2\varphi.
$$
The controlled map is given by Eq.~(\ref{eqn:Cmap}), which is
\begin{eqnarray*}
	&& A'=A-\partial_\varphi V(\varphi)-\partial_\varphi f_{\mathrm{Lie,a}} (\varphi),\\
	&&\varphi'=\varphi+\omega(A-\partial_\varphi V(\varphi)-\partial_\varphi f_{\mathrm{Lie,a}}(\varphi)),
\end{eqnarray*}
since $V$ and $f_{\mathrm{Lie,a}}$ do not depend on the action $A$. 
The associated controlled map is thus the same as the one obtained with mixed coordinates $\cS_{\mathrm{c,a}}$.

\subsection{Application to the tokamap}

\begin{figure}
\epsfig{file=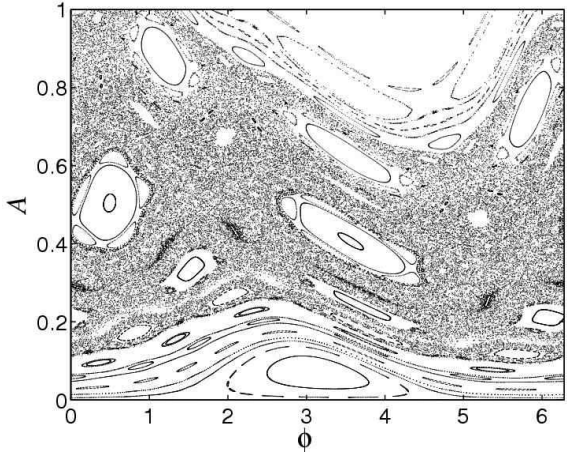,width=7.5cm,height=6.0cm}
\caption{Phase portrait of the tokamap for $\varepsilon=9/(4\pi)$.}
\label{fig:tok1}
\end{figure}

The tokamap~\cite{toka1,toka2} has been proposed as a model map for toroidal chaotic magnetic fields. It describes the motion of field lines on the poloidal section in the toroidal geometry. This symplectic map $(A,\varphi)\mapsto (A',\varphi')$, where $A$ is the toroidal flux and $\varphi$ is the poloidal angle, is generated by the function
$$
S(A',\varphi)=A'\varphi +H(A')-\varepsilon\frac{A'}{A'+1}w(\varphi),
$$  
with $w(\varphi) = \cos \varphi$. 
It reads
\begin{eqnarray*}
	&& A'=A-\varepsilon\frac{A'}{1+A'}\sin\varphi,\\
	&& \varphi'=\varphi+\frac{1}{q(A')}-\frac{\varepsilon}{(1+A')^2}\cos\varphi,
\end{eqnarray*}
where $q(A)=1/\omega(A)$ is called the $q$-profile. 
In our computation, we choose $\omega(A)\equiv \partial_A H=1/q(A)=\pi(2-A)(2-2A+A^2)/2$ and $\varepsilon=9/(4\pi)$. A phase portrait of this map is shown in Fig.~\ref{fig:tok1}.

The controlled map must satisfy some specific requirements~: In addition to being area-preserving, the toroidal geometry must be preserved which means that $A$ has to be positive for all times (for a circular section, $A=r^2$ where $r$ is the dimensionless radial coordinate).

The generating function determined by Eq.~(\ref{eqn:chi}) is
$$
\chi(A,\varphi)=\frac{\varepsilon A}{2(1+A)}\frac{\sin(\varphi-\omega(A)/2)}{\sin(\omega(A)/2)}.
$$
The control term given by Eq.~(\ref{eqn:hmix}) is
$$
f(A,\varphi)=\frac{\varepsilon^2 A}{2(1+A)^2}\frac{\cos(\varphi+\omega(A)/2)}{\sin(\omega(A)/2)}
\left[ \frac{\cos\varphi}{1+A}-A\partial_A \omega\frac{\cos(\varphi+\omega(A)/2)}{4\sin(\omega(A)/2)}\right].
$$

\begin{figure}
\epsfig{file=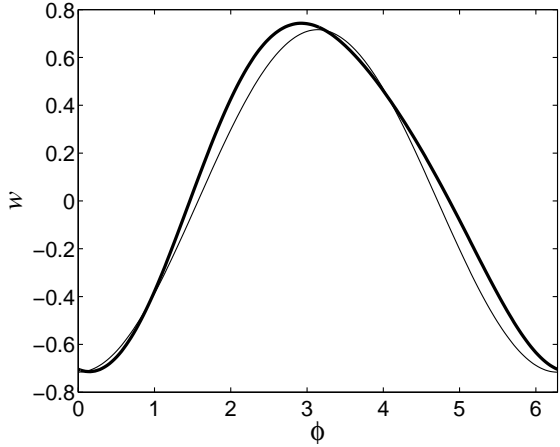,width=7.5cm,height=6.0cm}
\caption{Potential of the generating function of the tokamap with $\varepsilon=9/(4\pi)$~: $w$ without control (thin line) and $w_{\mathrm a}$ with control (thick line).}
\label{fig:Wtoka}
\end{figure}

\begin{figure}
\epsfig{file=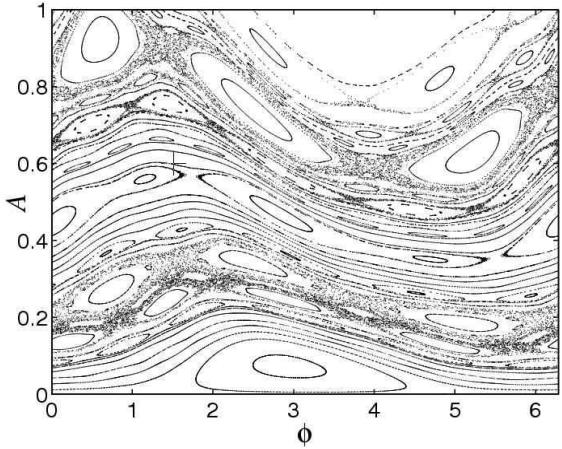,width=7.5cm,height=6.0cm}
\caption{Phase portrait of the controlled tokamap with the control term given by Eq.~(\ref{eqn:ctTok}) for $\varepsilon=9/(4\pi)$.}
\label{fig:tok2}
\end{figure}

We simplify the control term by considering the region near $A=1/2$. To preserve positive values for $A$, we keep the term $A/(1+A)$ in front of the control term. The control term becomes
\begin{equation}
\label{eqn:ctTok}
f_{\mathrm{a}}(A,\varphi)=\frac{\varepsilon^2 A}{1+A}\frac{\cos(\varphi+\alpha)}{3\sin\alpha}\left( \frac{2}{3}\cos\varphi +\frac{11\pi}{64}\frac{\cos(\varphi+\alpha)}{\sin\alpha}\right),
\end{equation} 
where $\alpha=15\pi/32$. It means that the potential $w$ of the tokamap is now slightly modified into
$$
w_{\mathrm{a}}(\varphi)=\cos\varphi+\varepsilon\frac{\cos(\varphi+\alpha)}{3\sin\alpha}\left( \frac{2}{3}\cos\varphi +\frac{11\pi}{64}\frac{\cos(\varphi+\alpha)}{\sin\alpha}\right).
$$
The functions $w$ and $w_{\mathrm{a}}$ are plotted on Fig.~\ref{fig:Wtoka}.
A phase portrait of the tokamap for $\varepsilon=9/(4\pi)$ with the control term given by Eq.~(\ref{eqn:ctTok}) is shown on Fig.~\ref{fig:tok2}. We clearly see that a lot of invariant tori have been created by the addition of the control term in the region $A=1/2$. This shows that a small and appropriate modification of the map (through the potential function) can drastically reduce the chaotic zones and hence the transport properties of the system.

%=========================================================== 
\begin{ack}
%=========================================================== 

We acknowledge useful discussions with D. Constantineanu, F. Doveil, D. Escande, R. Lima, U. Locatelli, J.H. Misguich, E. Petrisor, G. Steinbrecher and B. Weyssow. We acknowledge the financial support from Euratom/CEA 
(contract $V3382.001$), from the italian I.N.F.N. and I.N.F.M. 
G.C. thanks I.N.F.M. for financial support through a PhD fellowship. The work of Y.E. is partly supported by a delegation position from Universit\'e de Provence to CNRS.
\end{ack}

\appendix

%=========================================================== 
\section{Proof of ${\mathrm e}^{\{U\}}W=U\boxplus W\boxplus (-U)$}
\label{app:opl}
%=========================================================== 
First we recall the identity $\rme^{\{\alpha\}} \beta =
\rme^\alpha \beta \rme^{-\alpha}$ for linear operators $\alpha$ and $\beta$, where $\{\cdot \}$ is the commutator, i.e.\ $\{\alpha\} \beta =\alpha \beta -\beta \alpha$. We apply
it to $\alpha = \{U\}$ and $\beta = \{ W\}$ in $L(\cA)$, where $U$
and $W$ are functions in $\cA$. Then $\{\{ U\}\}$ is the linear
operator acting on $L(\cA)$ as the commutation with the element
$\{U\}$, i.e.\  $\{\{ U\}\}\{W\} =\{U\}\{W\}-\{W\}\{U\}$. Thus
$$
  \rme^{\{\{U\}\}} \{W\}
  =
  \rme^{\{U\}} \{W\} \rme^{-\{U\}} .
$$
On the other hand, we have
$$
  \exp (\rme^{\{U\}} \{W\} \rme^{-\{U\}})
  =
  \rme^{\{U\}} \rme^{\{W\}} \rme^{-\{U\}}
  =
  \rme^{\{U\} \oplus \{W\} \oplus \{-U\}},
$$
where the second equality follows from
definition of the warped addition $\oplus$ given by Eq.~(\ref{eqn:defoplus}).

The Jacobi identity implies
$\{\{U\}W\}=\{U\}\{W\}-\{W\}\{U\}=\{\{U\}\}\{W\}$, from which it
follows by recursion that $\{\{U\}^n W\}=\{\{U\}\}^n\{W\}$, for
all $n\in {\mathbb Z}$, and hence
$$
  {\mathrm e}^{\{\{U\}\}}\{W\}
  =
  \{{\mathrm e}^{\{U\} } W\}.
$$
Since we have $\{U\}\oplus\{ W\}=\{U\boxplus
W\}$ by definition~(\ref{eqn:defboxplus}), we obtain Eq.~(\ref{eqn:lem}).

%===========================================================
\section{Expression of the generating function $S_0$}
\label{app:1}
%===========================================================
In what follows, the functions $V_A$ and $V_\varphi$ denote the partial derivatives of any function $V$ with respect to $A$ and $\varphi$. In the same way, $V_{A\varphi}$ denotes the second partial derivative $\partial_A\partial_\varphi V$. The determination of $S_0$ follows the scheme depicted on Fig.~\ref{fig:map}. The computation is done mainly in two steps~: First, we derive the generating function which maps the variables $(A,\varphi)$ to $(A'_0,\varphi'_0)$ by eliminating the dependency on the variables $(A',\varphi')$. Secondly, we derive the expression of the generating function $S_0$ which maps the variables $(A_0,\varphi_0)$ to $(A'_0,\varphi'_0)$ by eliminating the dependency on the variables $(A,\varphi)$.

The variables $(A'_0,\varphi'_0)$ are defined implicitly as functions of $(A',\varphi')$ by the canonical transformation generated by $\chi$~:
\begin{eqnarray}
    && A'=A'_0+\varepsilon \chi_\varphi (A'_0,\varphi'),\label{eqn:YE1}\\
    && \varphi'_0=\varphi'+\varepsilon \chi_A(A'_0,\varphi').\label{eqn:YE2}
\end{eqnarray}
The generating function $S_{\mathrm{c}}$ gives
\begin{eqnarray}
    && A=A'+\varepsilon V_\varphi(A',\varphi)+\varepsilon^2 f_\varphi(A',\varphi),\label{eqn:YE3}\\
    && \varphi'=\varphi+\omega(A')+\varepsilon V_A(A',\varphi)+\varepsilon^2 f_A(A',\varphi).\label{eqn:YE4}
\end{eqnarray}
First, we rewrite Eq.~(\ref{eqn:YE4}) by using Eq.~(\ref{eqn:YE1}) in order to get the expression of $\varphi'$ in terms of the variables $(A'_0,\varphi)$ up to order $\varepsilon^2$~:
\begin{eqnarray*}
    \varphi'&=&\varphi+\omega(A'_0)+\varepsilon \left[ \omega_A(A'_0)\chi_\varphi (A'_0,\varphi+\omega(A'_0))+V_A(A'_0,\varphi)\right]\\
    &&+\varepsilon^2\Bigl[ \omega_A(A'_0)^2 \chi_\varphi (A'_0,\varphi+\omega(A'_0))\chi_{\varphi\varphi}(A'_0,\varphi+\omega(A'_0))\\
    && \quad +\omega_A(A'_0)V_A(A'_0,\varphi)\chi_{\varphi\varphi}(A'_0,\varphi+\omega(A'_0))+\frac{1}{2}\omega_{AA}(A'_0)\chi_\varphi^2(A'_0,\varphi+\omega(A'_0))\\
    && \quad  +V_{AA}(A'_0,\varphi) \chi_\varphi (A'_0,\varphi+\omega(A'_0))+f_A(A'_0,\varphi)\Bigr]+O(\varepsilon^3).
\end{eqnarray*}
We substitute the expression of $\varphi'$ into Eq.~(\ref{eqn:YE2})~:
\begin{eqnarray*}
    \varphi'_0&=&\varphi+\omega(A'_0)+\varepsilon \left[ \omega_A(A'_0)\chi_\varphi (A'_0,\varphi+\omega(A'_0))+\chi_A (A'_0,\varphi+\omega(A'_0))+V_A(A'_0,\varphi)\right]\\
    &&+\varepsilon^2\Bigl[ \omega_A(A'_0)^2 \chi_\varphi (A'_0,\varphi+\omega(A'_0))\chi_{\varphi\varphi}(A'_0,\varphi+\omega(A'_0)) \\
    && \quad +\omega_A(A'_0)V_A(A'_0,\varphi)\chi_{\varphi\varphi}(A'_0,\varphi+\omega(A'_0))+\frac{1}{2}\omega_{AA}(A'_0)\chi_\varphi^2(A'_0,\varphi+\omega(A'_0))\\
    && \quad +V_{AA}(A'_0,\varphi) \chi_\varphi (A'_0,\varphi+\omega(A'_0))+V_A(A'_0,\varphi)\chi_{A\varphi}((A'_0,\varphi+\omega(A'_0))\\
    && \quad  +  \omega_A(A'_0)\chi_\varphi(A'_0,\varphi+\omega(A'_0))\chi_{A\varphi}(A'_0,\varphi+\omega(A'_0))
    +f_A(A'_0,\varphi)\Bigr]+O(\varepsilon^3).
\end{eqnarray*}
The substitution of the expression of $\varphi'$ into Eq.~(\ref{eqn:YE1}) gives $A'$ as a function of $A'_0$ and $\varphi$. Inserting this expression into Eq.~(\ref{eqn:YE3}) gives
\begin{eqnarray*}
    A&=&A'_0+\varepsilon\left[ \chi_\varphi(A'_0,\varphi+\omega(A'_0))+V_\varphi(A'_0,\varphi)\right]\\
    && +\varepsilon^2\left[ V_{A\varphi}(A'_0,\varphi)\chi_\varphi (A'_0,\varphi+\omega(A'_0))
    +V_{A}(A'_0,\varphi)\chi_{\varphi\varphi} (A'_0,\varphi+\omega(A'_0))\right.\\
    &&\left.  \quad +\omega_A(A'_0)\chi_\varphi (A'_0,\varphi+\omega(A'_0))\chi_{\varphi\varphi}(A'_0,\varphi+\omega(A'_0))+f_\varphi(A'_0,\varphi)\right]+O(\varepsilon^3).
\end{eqnarray*}
The expressions of $A$ and $\varphi'_0$ as functions of $A'_0$ and $\varphi$ can be obtained by the generating function~:
\begin{eqnarray*}
\tilde{S}_0(A'_0,\varphi)&=&A'_0\varphi +H(A'_0)+\varepsilon\left[ V(A'_0,\varphi)+\chi(A'_0,\varphi+\omega(A'_0))\right]\\
&&+\varepsilon^2\Bigl[ V_A(A'_0,\varphi) \chi_\varphi(A'_0,\varphi+\omega(A'_0))\\ && \quad  +\frac{1}{2}\omega_A(A'_0) \chi^2_\varphi(A'_0,\varphi+\omega(A'_0)) +f(A'_0,\varphi) \Bigr] +O(\varepsilon^3).
\end{eqnarray*}
Next, we eliminate the variables $A$ and $\varphi$ using the equations~:
\begin{eqnarray}
    && A=A_0+\varepsilon \chi_\varphi (A_0,\varphi),\label{eqn:YE11}\\
    && \varphi_0=\varphi+\varepsilon \chi_A (A_0,\varphi).\label{eqn:YE12}
\end{eqnarray}
In order to perform the same type of procedure to eliminate $A$ and $\varphi$, we first need to invert Eqs.~(\ref{eqn:YE11})-(\ref{eqn:YE12})~:
\begin{eqnarray*}
    && A_0=A-\varepsilon \chi_\varphi (A,\varphi_0)\\
    && \qquad +\varepsilon^2\left[ \chi_\varphi (A,\varphi_0)\chi_{A\varphi} (A,\varphi_0)
    +\chi_A (A,\varphi_0)\chi_{\varphi\varphi} (A,\varphi_0)\right]+O(\varepsilon^3),\\
    && \varphi=\varphi_0-\varepsilon \chi_A (A,\varphi_0)\\
    && \qquad +\varepsilon^2\left[ \chi_A (A,\varphi_0)\chi_{A\varphi} (A,\varphi_0)
    +\chi_\varphi (A,\varphi_0)\chi_{AA} (A,\varphi_0)\right]+O(\varepsilon^3),
\end{eqnarray*}
which can also be obtained from the generating function
$$
\tilde{X}(A,\varphi_0)=A\varphi_0-\varepsilon \chi(A,\varphi_0)+\varepsilon^2 \chi_{A} (A,\varphi_0)\chi_{\varphi} (A,\varphi_0)+O(\varepsilon^3).
$$
The same type of substitutions as for eliminating $(A',\varphi')$ leads to the expression of $A_0$ and $\varphi'_0$ as functions of $A'_0$ and $\varphi_0$~:
\begin{eqnarray*}
    && A_0=A'_0+\varepsilon\left[ \chi_\varphi (A'_0,\varphi_0+\omega(A'_0))-\chi_\varphi(A'_0,\varphi_0) +V_\varphi(A'_0,\varphi_0)\right]\\
    && \quad +\varepsilon^2 \Bigl[ V_{A\varphi}(A'_0,\varphi_0)\chi_\varphi (A'_0,\varphi_0+\omega(A'_0))+V_A (A'_0,\varphi_0)\chi_{\varphi\varphi}(A'_0,\varphi_0+\omega(A'_0)) \\
    && \qquad -V_{\varphi\varphi}(A'_0,\varphi_0)\chi_A(A'_0,\varphi_0)-\chi_{\varphi\varphi}(A'_0,\varphi_0+\omega(A'_0))\chi_A(A'_0,\varphi_0)\\
    && \qquad +\omega_A(A'_0)\chi_\varphi(A'_0,\varphi_0+\omega(A'_0))\chi_{\varphi\varphi}(A'_0,\varphi_0+\omega(A'_0))\\
    && \qquad + \chi_{A\varphi}(A'_0,\varphi_0)(\chi_\varphi(A'_0,\varphi_0)-\chi_\varphi(A'_0,\varphi_0+\omega(A'_0)))\\
    && \qquad  +\chi_{\varphi\varphi}(A'_0,\varphi_0)\chi_A(A'_0,\varphi_0)-V_\varphi(A'_0,\varphi_0)\chi_{A\varphi}(A'_0,\varphi_0) +f_\varphi(A'_0,\varphi_0)   \Bigr] +O(\varepsilon^3),\\
    && \varphi'_0=\varphi_0+\omega(A'_0)+\varepsilon\left[V_A (A'_0,\varphi_0)+\omega_A(A'_0) \chi_\varphi(A'_0,\varphi_0+\omega(A'_0))\right.\\
    && \qquad \left. +\chi_A(A'_0,\varphi_0+\omega(A'_0))-\chi_A(A'_0,\varphi_0)\right]\\
    && \quad +\varepsilon^2 \Bigl[ \chi_A(A'_0,\varphi_0) \chi_{A\varphi}(A'_0,\varphi_0) +\chi_{\varphi}(A'_0,\varphi_0) \chi_{AA}(A'_0,\varphi_0) \\
    && \qquad -\chi_A(A'_0,\varphi_0) V_{A\varphi}(A'_0,\varphi_0) -\omega_A(A'_0) \chi_A(A'_0,\varphi_0) \chi_{\varphi\varphi}(A'_0,\varphi_0+\omega(A'_0)) \\
    && \qquad -\chi_A(A'_0,\varphi_0) \chi_{A\varphi}(A'_0,\varphi_0+\omega(A'_0))  + \chi_\varphi(A'_0,\varphi_0+\omega(A'_0)) V_{AA}(A'_0,\varphi_0) \\
    && \qquad +V_A(A'_0,\varphi_0) \left( \chi_{A\varphi}(A'_0,\varphi_0+\omega(A'_0)) +\omega_A(A'_0) \chi_{\varphi\varphi}(A'_0,\varphi_0+\omega(A'_0)) \right)\\
    && \qquad +\omega_A(A'_0) \chi_\varphi(A'_0,\varphi_0+\omega(A'_0)) \chi_{A\varphi}(A'_0,\varphi_0+\omega(A'_0))\\
    && \qquad +\omega_A(A'_0)^2 \chi_\varphi(A'_0,\varphi_0+\omega(A'_0)) \chi_{\varphi\varphi}(A'_0,\varphi_0+\omega(A'_0))\\
    && \qquad - \chi_{AA}(A'_0,\varphi_0) \left( \chi_\varphi(A'_0,\varphi_0+\omega(A'_0)) +V_\varphi(A'_0,\varphi_0) \right)\\
    && \qquad  +\frac{1}{2}\omega_{AA}(A'_0) \chi_\varphi^2(A'_0,\varphi_0+\omega(A'_0)) +f_A(A'_0,\varphi_0)  \Bigr] +O(\varepsilon^3).
\end{eqnarray*}
The expressions of $A_0$ and $\varphi'_0$ can be obtained using the generating function
\begin{eqnarray*}
    S_0(A'_0,\varphi_0)&=& A'_0\varphi_0+H(A'_0)+\varepsilon\left[ V(A'_0,\varphi_0)+\chi(A'_0,\varphi_0+\omega(A'_0))-\chi(A'_0,\varphi_0)\right]\\
    && +\varepsilon^2\Bigl[ \left(V_A(A'_0,\varphi_0) -\chi_A(A'_0,\varphi_0) \right) \chi_\varphi(A'_0,\varphi_0+\omega(A'_0)) \\
    && \quad  - \left(V_\varphi(A'_0,\varphi_0) -\chi_\varphi(A'_0,\varphi_0) \right) \chi_A(A'_0,\varphi_0)  \\
    && \quad + \frac{1}{2}\omega_A(A'_0)\chi_\varphi^2(A'_0,\varphi_0+\omega(A'_0))+f(A'_0,\varphi_0) \Bigr]+O(\varepsilon^3).
\end{eqnarray*}
Using Eq.~(\ref{eqn:defz}) yields Eq.~(\ref{eqn:s0}).

\end{document}